\documentclass[pre,twocolumn,unsortedaddress,floatfix,amssymb]{revtex4} 

\usepackage{times}
\usepackage{graphicx}
\usepackage{amsmath}
\DeclareGraphicsExtensions{.eps}

\begin{document}

\title{Critical evaluation of the computational methods used in the forced polymer translocation}

\author{V. V. Lehtola}
\author{R. P. Linna}
\author{K. Kaski}
\affiliation{
  Department of Biomedical Engineering and Computational Science,
  Helsinki University of Technology,
  P.O. Box 9203, FIN-02150 TKK, Finland}

\date{June 1, 2008}

\begin{abstract}
In forced polymer translocation, the average translocation time, $\tau$,
scales with respect to pore force, $f$, and polymer length, $N$, as $\tau
\sim f^{-1} N^{\beta}$. We demonstrate that an artifact in Metropolis Monte
Carlo method resulting in breakage of the force scaling with large $f$ may
be responsible for some of the controversies between different
computationally obtained results and also between computational and
experimental results. Using Langevin dynamics simulations we show that the
scaling exponent $\beta \le 1 + \nu$ is not universal, but depends on $f$.
Moreover, we show that forced translocation can be described by a relatively
simple force balance argument and $\beta$ to arise solely from the initial
polymer configuration. 
\end{abstract}

\maketitle

\section{Introduction}

\begin{figure}
\centerline{
\includegraphics[angle=0, width=0.5\textwidth]{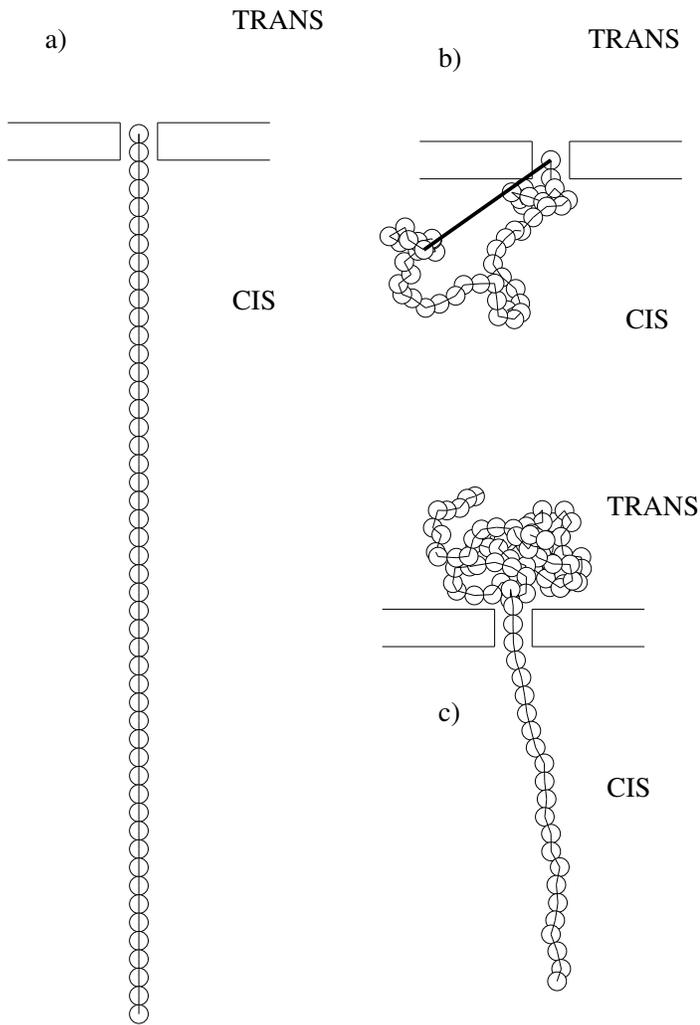}
}
\caption{
Snapshots from 3D Langevin dynamics simulations.
a) A straight initial configuration of a chain with $N=50$ beads.
b) A relaxed equilibrium initial configuration of a chain with $N=50$ beads.
The wide solid line is the pore-to-end distance $R_{pe}$.
c) A chain with $N=100$ beads translocating with a force of $f/D_0 =
2.91$, when $s=80$ beads have translocated.
}
\label{straight_conf}
\end{figure}

The force-driven transport of biopolymers through a nano-scale
pore in a membrane is a ubiquitous process in biology. Despite the
complex dynamics involved in the process, Monte Carlo (MC) has been
almost the only computational method used for modeling it, see the schematic
Fig.~\ref{straight_conf}. Only fairly recently more realistic dynamics have been
applied~\cite{ali,fyta,bernaschi,gauthier,lehtola}. The classic
theoretical treatment based on writing down the free energy for a
system of two equilibrium ensembles separated by a
wall~\cite{sung,muthukumar}, henceforth called Brownian translocation, has not
found support from MC simulations. The close-to-equilibrium dynamics
would validate the derivation of the translocation dynamics using the
Rouse relaxation time or diffusion based arguments. The
validity of assumptions about the diffusion constant along the polymer
chain was questioned already by the authors of~\cite{chuang,kantor}, who also
noted that the characteristic translocation time both for phantom and
self-avoiding polymer chains was bound to be greater than their
characteristic time for relaxation to thermal equilibrium.  The
theoretical treatment of forced translocation can be said to have so
far been almost solely guided by MC
simulations~\cite{dubbeldam_pre,dubbeldam_epl}, despite the MC results
contradicting the available experimental results. Hence, the theory has evolved
independently of the experimental findings.

In the attempt to determine the dynamical universality of
translocation processes the scaling of the average translocation time
$\tau$ with respect to the polymer chain length $N$ as $\tau \sim
N^\beta$ has been under intensive study. There is an abundance
of research reporting different scaling exponents, $\beta$. Some of the
lately reported results do find some consistency on $\beta$. For
example, both the MC and Langevin dynamics (LD) simulations were reported
to give $\beta = 2 \nu$ for short and $\beta = 1+ \nu$ for long polymers in
2D, where $\nu$ is the swelling
exponent~\cite{luo_mc,luo_langevin}. Strictly universal scaling was
claimed in~\cite{dubbeldam_epl}, where $\beta = 1.5$ was obtained both
in 2D and 3D.

In addition, several computational investigations have suggested
nearly linear scaling ($\beta = 1$). On one hand, this close-to-linear
behavior is prone to appear at the over-damped ({\it i.e.} Brownian)
limit without self-avoiding effects, as shown by MC
simulations in~\cite{gauthier_jcp}. On the other hand, in~\cite{tian} $\beta$
value close to unity could be inferred, but the polymer chains were
found to be clearly out of equilibrium both on {\it cis} and {\it trans} sides
in Brownian dynamics simulations, which was ignored in the
interpretation of the results. Linear scaling has also been obtained
from MC simulations in~\cite{loebl} and from an LD
simulation~\cite{forrey}, albeit for moderate ranges of $N$.

Not only do the different computational results and accompanying theories
contradict, but they also seem to have very little to offer in explaining the
available experimental results. For example, the experimental
study by Storm {\it et al.}~\cite{storm} reports $\beta = 1.27$,
whereas the FFE framework~\cite{dubbeldam_pre,dubbeldam_epl} claims
uniform scaling different from this at all forces and is thus
obviously in trouble here. Since there logically exists a requirement for
the results obtained by using more realistic computational methods to
be validated not only against experimental results but also against
results from MC model simulations, a demand is placed for
identifying the limitations and possible errors of the MC based
translocation models. 

Our recent study using molecular-dynamics-based simulations showed
that $\beta$ depends on the pore force and that the obtained values for
$\beta$ are in accord with the experiments~\cite{lehtola}. As pointed
out by Storm {\it et al}~\cite{storm}, the experimental pore force
magnitudes are anticipated to be larger than those used in computer
simulations, which would explain the differing $\beta$ values. However, larger
force magnitudes in MC have been claimed to produce the same scaling
relations as smaller ones. We suggest that saturation of transition
probabilities for large forces in MC could be responsible for this discrepancy
between simulated and experimental $\beta$
values~\cite{kasianowicz,storm,meller,li,chen,huang}.

Accordingly, we set out to study the forced polymer translocation with the
pore force, $f$, as a control parameter. Three different methods will
be used for modeling the process. The standard MC using Metropolis
sampling will be compared with kinetic (or n-fold) Monte Carlo method
(KMC)~\cite{young, bortz} in 1D. This comparison is made in order to
characterize the forced translocation process. Differences between MC
and KMC results would indicate subtleties in the effective
independence of the events comprising the
system~\cite{fichthorn}. LD simulations will be used as
a reference in 1D. With LD we also compute results in 2D and 3D to
gain further evidence for our description of
the forced polymer translocation. Finally, we determine the
effect of the initial configuration on the scaling of the translocation
time, $\tau$. In order to be free from potential artifacts due to
spatial discretization, all results are produced by off-lattice simulations. 

\section{Background}
The investigation of the dynamics of forced translocation is subtly related
to the assumption of close-to-equilibrium dynamics.  The classic treatments
of the problem by Sung and Park~\cite{sung} and Muthukumar~\cite{muthukumar}
assumed that a free energy, $F_s$, could be written down for the two
ensembles on either side of the wall, when $s$ segments have been
translocated:
\begin{equation}
\frac{F_s}{k_B T} = (1- \gamma) \ln \big[ s(N - s) \big] - \frac{s
  \tilde{f} \Delta z}{k_B T},
\label{free_energy}
\end{equation}
where $\tilde{f}$ is the pore force, and $\gamma = 0.69$ for a
self-avoiding chain~\cite{muthukumar}.
By using the free energy one can write down the one-dimensional Langevin
equation for translocation as a function of translocation coordinate,
$s$, when entropic effects are assumed small:
\begin{equation}
m \dot{s} = - m \xi \partial{F_s}/\partial{s} + \eta(t) = - m \xi \tilde{f} \Delta z + \eta(t),
\label{langevin}
\end{equation}
where $m$, $\xi$, $\eta(t)$, and $\Delta z$ are the bead mass, friction
constant, random force, and pore length, respectively.  From the Langevin
equation the scaling of the translocation time with the pore force $\tau
\sim f^{-1}$ is straightforwardly obtained, where $f \equiv \tilde{f} b / k_B T$ and
$b$ are the dimensionless pore force and the Kuhn length, respectively.
Although the form of the Langevin equation (Eq. (\ref{langevin})) is identical
in forced and unforced translocation, the connection between the free
energy, $F_s$, and the pore force, $\tilde{f}$, breaks down in forced translocation,
where the pore force is a control parameter. The Langevin equation approach
yields $\tau \propto N^2$ and $\tau \propto N^1$ as bounds for the
translocation time scaling~\cite{muthukumar,kantor}. Based on mere
unhindered motion of polymers over potential difference $\Delta \mu \sim f$
and the initial equilibrium configuration, Kantor and Kardar predicted that
the translocation time should scale as~\cite{kantor}

\begin{equation}
\tau \sim \frac{N^{1+\nu}}{f}.
\label{kant_scale}
\end{equation}

By the computational method where polymer follows detailed molecular
dynamics and the solvent coarse grained stochastic rotation dynamics, 
we have previously found that the translocating polymers are driven
increasingly out of equilibrium on both sides of the pore under
pore force magnitudes relevant in biological and experimental
systems~\cite{lehtola}. The dynamics is
then mainly determined by the force balance between the drag
force exerted on the mobile beads on the {\it cis} side and the
constant pore force, $f_d = f$. Additional contribution comes from the
crowding of the polymer beads on the {\it trans} side due to relaxation towards
equilibrium being slower than the rate at which new segments
enter through the pore. On the {\it cis} side the rate at which
polymer beads are set in motion towards the pore was seen to be
greater than the rate at which beads entered the pore, which indicates
that the tension spreads along the polymer faster than what the
polymer is able to relax towards equilibrium. By measuring the rate
at which the tension spreads, an estimate could then be made for the
scaling of the translocation time with the polymer length. We will
present these observations in the Results section obtained from
LD that, due to the absence of hydrodynamics, shows these
characteristics even more clearly.

\section{Polymer Model}
In the model system adjacent
monomers are connected with anharmonic springs, described by the
finitely extensible nonlinear elastic (FENE) potential,
\begin{equation}
U_{FENE} = - \frac{K}{2} R^2 \ln \big ( 1- \frac{(l - l_0)^2}{R^2} \big ).
\label{fene_mc}
\end{equation}
Here $l$ is the length of an effective bond, which can vary between
$l_{min} < l < l_{max}$,  $R = l_{max} - l_0 = l_0 - l_{min}$, and $l_0$
the equilibrium distance at which the bond potential takes its minimum
value. Choosing $l_{max}$ as the unit length and $R = 0.3$, yields
$l_{min} = 0.4$, and $l_0 = 0.7$. In the standard MC
simulations $k_B T = 1$ and the spring constant $K = 40\ k_B
T$. KMC dynamics proved more susceptible to bond fluctuations than
standard MC, so $k_B T = 0.1$ and $K/k_B T = 400$ were
used (see next section). The FENE potential suffices in $1D$. 

For 2D and 3D, where LD was used, the Lennard-Jones (LJ) potential,
\begin{eqnarray}
U_{LJ} &=& 4 \epsilon \left[ \left(\frac{\sigma}{r}\right)^{12} - \left(\frac{\sigma}{r}\right)^{6} \right]
, \: r \leq 2^{-1/6} \sigma \\
U_{LJ} &=& 0 , \; r > 2^{-1/6} \sigma,
\label{lj_md}
\end{eqnarray}
was used between all and FENE-potential, Eq. (\ref{fene_mc}), between adjacent
beads. The parameter values were chosen to be  $\epsilon = 1.2$, and
$\sigma = 1.0$ for the LJ, and $l_0 = 0$, $R = 1.5 = l_{max}$, $K = 60
/ \sigma^2$ for the FENE potential. The used LJ potential with no
attractive part mimics good solvent condition for the
polymers. Initial states are relaxed equilibrium configurations.

\section{Translocation Models}
In 1D we study the translocation dynamics using MC, KMC, and
LD methods. In order to link the 1D dynamics with forced
translocation we define the segment $s$ as translocated, when it
passes the original position of the first bead. Time units cannot be
expected to be identical for the three methods, but the scaling laws
given by MC and KMC must be identical to those obtained from LD, if
the MC and KMC methods are to be taken as representative
models for forced translocation.

One-dimensional off-lattice Monte Carlo simulations are performed with
the standard Metropolis acceptance test. A transition
to a new state is attempted by picking at random a bead, computing
the change in the particle's potential energy resulting from an
attempted trial move by a distance $\delta$ from its present position
$r$, $\Delta U = U(r+\delta) -U(r)$. Transition is always accepted for
$\Delta U < 0$, and according to its Boltzmann weight $\exp(-\beta
\Delta U)$, if $\Delta U > 0$. Time is incremented after each attempt, whether
accepted or rejected, by a constant amount $\Delta t$. For a detailed
pseudocode, see Appendix A.

Unlike in the Metropolis MC method, in KMC algorithm the system is
moved to another state at every attempt, regardless of how
improbable the transition is. Accordingly, KMC has been the
choice for doing simulations at low temperatures, where transition
probabilities are low and, accordingly, Metropolis MC
prohibitively slow. On the other hand, the additional bookkeeping
required in KMC makes it computationally slower than MC at higher
temperatures. However, MC and KMC have profound differences in other
aspects than just computational efficiency. 

In KMC the probability of the move is reflected on the (stochastically)
estimated time, $\Delta t$, for the transition to take place, see Appendix
B. Estimation of the elapsed time involves computing the system's all transfer
probabilities, or rates, $p_j$, from which a cumulative function $R_i
= \sum_{j=1}^i p_j$ is calculated. $R_i$ is a class including events,
whose probabilities $p_j \in (p^{min}_i,p^{max}_i]$. From the possible
transitions $j$ the event to take place is picked at random, so that
$R_{i-1} < uR(t) < R_i$, where $u \in (0,1]$ is a random number and
$R(t) = R_l$ and $l$ the number of transition classes. The time
elapsed between the previous and current event is estimated as $\Delta
t = - 1/R(t) \ln u'$, where $u' \ne u$ is a random number. It is
noteworthy that the stochastically determined time increment $\Delta
t$ is inversely proportional to the the total probability, {\it i.e.}
the rate of change, $R(t)$, of the system evolving in time, here the
instantaneous polymer configuration.

Due to finite distance for the trial move, $\delta$, the Metropolis MC
inevitably eliminates very improbable moves, which in the case of our
1D polymer simulations are those stretching or compressing polymer
bonds far from their equilibrium lengths, $l_0$. Hence, as seen in our
simulations, bond length
fluctuations are larger in KMC than in MC. To maintain stability we
used a lower temperature in our KMC simulations.  A mere
scaling of the bond potential magnitude by reducing the temperature
does not change the dynamic universality class of the system.

Our LD algorithm was implemented according to~\cite{allen}.
The LD translocation algorithm was used also in two and three
dimensions. In the algorithm, the pore force, $f$, is a free control
parameter and not derived from free energy, as was done in
Eq.~(\ref{langevin}). In the initial states the bond lengths are equal
to relaxation distances and the first bead is in the middle of the
pore. The pore diameter is $1.2 b$ and length $3b$, where $b=1$ is the
Kuhn length of the modeled polymer. We have used $k_B T = 1$, $\xi =
0.73$, and $m = 16$ parameters in the Langevin implementation. This
yields $D_0 \equiv k_B T/ \xi m \approx 0.086$ for the one-particle
self-diffusion constant in 1D.

\section{Results}

Our motivation for evaluating MC based methods in the context of forced
translocation comes on one hand from discrepancies between scaling
relations $\tau \sim N^\beta$ obtained from different MC simulations and
on the other hand discrepancy between all MC simulations and experiments.
We also look into the observed separate scaling regimes for short
and long polymers. We base our evaluation of
the methods mainly on the scaling of the translocation time with pore
force and with polymer length. Since we take LD as the reference for
the physical translocation system when
hydrodynamic interactions are ignored, we first check these scaling
relations using LD.

\begin{figure*}
\centerline{
\includegraphics[angle=0, height=0.20\textheight]{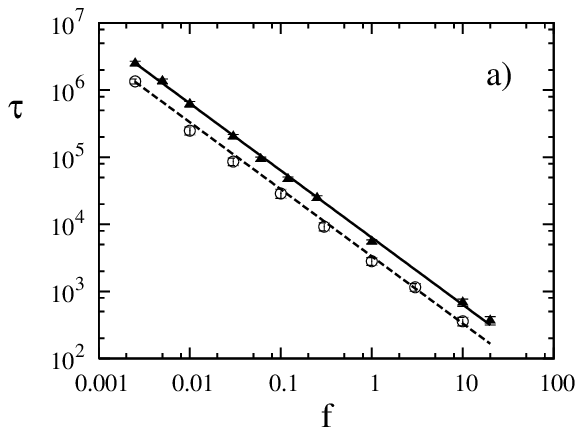}
\includegraphics[angle=0, height=0.20\textheight]{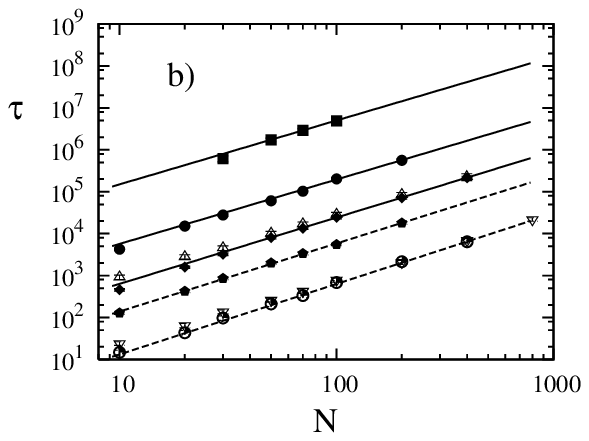}
\includegraphics[angle=0, height=0.20\textheight]{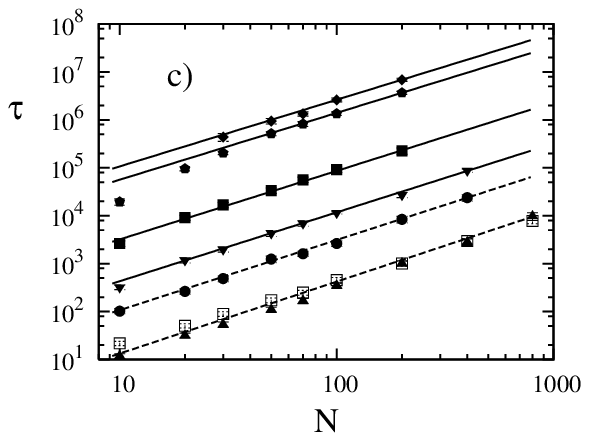}
}
\caption[MD force scaling] {
a) The scaling of the translocation time $\tau$ with respect to the external
force $f$.  Results from 2D ($\blacktriangle$) and 3D ($\circ$) Langevin dynamics
simulations yield the same $f \sim N^{-1}$ scaling. Since the simulations were
done with the only difference in dimension, we note that in higher dimension
the translocation process is faster. 
b) 2D and c) 3D Langevin dynamics simulation results for
the scaling of the translocation time $\tau$ with respect to chain
length $N$ with various forces. 
Open symbols are with frictional pore and full symbols are with a ballistic pore.
All plots yield different values for the scaling
exponent $\beta$ which are reported in Table~\ref{beta_values} along
with their corresponding forces.
The finite size effect due to the frictional is identical to $1D$ case
see text). With $f = 10$ only chains with $N \geq 200$ follow
the asymptotic scaling behavior. If the pore is non-aqueous (i.e. ballistic)
the finite size effect applies only for chains smaller than $N=30$, and is
amplified by reducing the force.
}
\label{scaling_moldy_fig}
\end{figure*}

Fig.~\ref{scaling_moldy_fig}~a) shows the scaling of the
translocation time with respect to the dimensionless pore force $ \tau \sim
f^\alpha$ obtained from LD simulations in 2D and 3D. The thermal
energy $k_B T$ was kept fixed in all simulations.
We obtained $\alpha = -0.990 \pm 0.01$ in 2D and
$\alpha = -0.978 \pm 0.02$ in 3D. So, the scaling in these
dimensions is the expected $\tau \sim f^{-1}$ in accordance with
Eq.~(\ref{langevin}), confirming that entropic effects are weak.

Fig.~\ref{scaling_moldy_fig} shows the scalings of the translocation
time with respect to chain length, $\tau \sim N^\beta$, for a
frictional and frictionless pore in 2D in b) and in 3D in
c). The scaling exponents $\beta$, identical for longer chains with frictional
and frictionless pores, are shown in Table~\ref{beta_values} for
different values of $f/D_0$. There is a fair agreement between our 2D
scaling exponents and those reported
previously~\cite{luo_langevin}. Qualitatively, the change of $\beta$
with $N$ with the frictional pore agrees with the aforementioned LD
results where two scaling regimes were claimed. Accordingly, we obtain
a smaller $\beta$ for shorter chains. However, a quite different
behavior is seen in the case of a frictionless pore. At a large enough
force there exists only one scaling identical to the common scaling
obtained for long polymers with both pores. Hence the translocation
times longer than what would be expected from this scaling seem to be
due to friction in the confined region inside the pore. 
\begin{table}[ht]
\caption{
Langevin dynamics. Values of $\beta$ obtained from Fig.~\ref{scaling_moldy_fig}.
Here $U = \tilde{f} \Delta z / k_B T$ is the dimensionless pore potential.
}
\begin{tabular}{cccc}
$U$ & $f/D_0$ & $\beta$ $(d=2)$ & $\beta$ $(d=3)$\\
\hline
0.0036 & 0.014 & $1.52 \pm 0.08$ & $1.40 \pm 0.04$ \\
0.0075 & 0.03 & - & $1.40 \pm 0.02$ \\
0.03   & 0.35 & $1.54 \pm 0.03$ & $1.43 \pm 0.02$ \\
0.75   & 2.91 & $1.58 \pm 0.01$ & $1.44 \pm 0.03$ \\
3      &  11    & $1.63 \pm 0.04$ & $1.47 \pm 0.05$ \\
30     & 116   & $1.68 \pm 0.02$ & $1.50 \pm 0.01$ \\
\end{tabular}
\label{beta_values}
\end{table}

The scaling exponents $\beta$ obtained in 2D, while agreeing with the results
in~\cite{luo_langevin}, seem to be in strong disagreement with the
scaling $\beta = 1.53$ obtained by 2D MC simulations for an infinite
potential with the pore length $\Delta z = 1$. In addition, in another
MC simulation in 2D  $\beta = 1.70$ has been obtained for two
different forces $f = 1, 5$ with $\Delta z = 3$~\cite{luo_mc}. So, not
only MC results contradict with LD results, but MC simulations
mutually disagree. To check if this could be due to MC translocation
model possibly belonging to a different dynamic universality class from the
respective LD model we simulated the simplest possible case, {\it
  i.e.} 1D forced translocation using KMC in addition to MC and LD. We
obtain invariably $\beta = 2$ for LD, MC, and KMC, in agreement with
the previous MC result~\cite{kantor}, so the computational MC model
using Metropolis sampling seems to produce correct time dependence,
although forced translocation can hardly  be taken  as a purely
Poissonian process~\cite{fichthorn}. However, we do find a
method-dependent artifact which explains the apparent discrepancy.

\begin{figure}[htbp]
\centerline{
\includegraphics[angle=0,height=0.13\textheight]{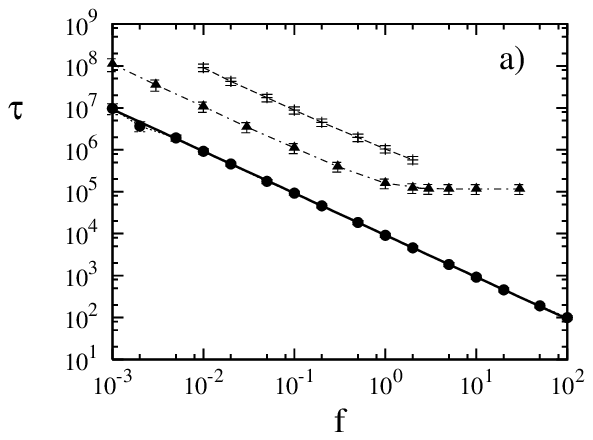}
\includegraphics[angle=0,height=0.13\textheight]{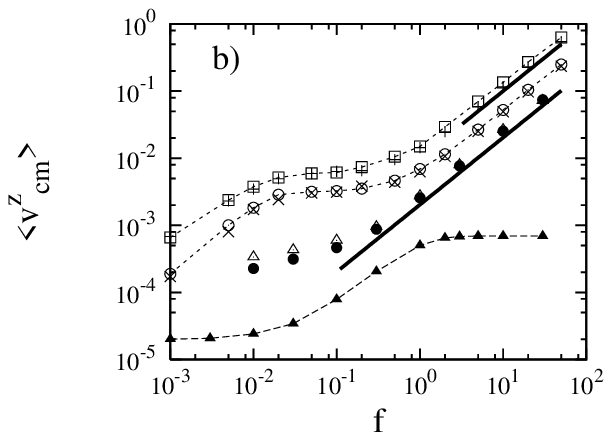}
} 
\caption {
a) Comparison of the scaling of translocation time $\tau$
(in arbitrary units) with pore force, $f$, in 1D obtained from MC
($\blacktriangle$), KMC ($\times$) and LD ($\bullet$) simulations. Polymers
are of constant length $N = 50$. All computational methods
yield the same force scaling $\tau \sim f^{-1}$ for $f \leq 1$. The MC
method shows an artifact in the force regime $f \gg 1$. KMC fails to
perform in this regime (see text).
b) The average center-of-mass velocity in the $z$-direction,
$\langle v_{cm}^z \rangle$, as a function of the
dimensionless pore force $f$, for $N=20$ ($+, \square$) and $N=50$
($\times, \circ$) in 1D, for $N=100$ ($\bullet$) in 2D, and for $N=100$
($\triangle$) in 3D. The data for ($\square$) and ($\circ$) was
obtained using a frictionless pore in LD simulations. All other
results are from LD simulations using a frictional pore.  Linear
scaling is plotted as a solid line to guide the eye.
At the bottom, results from 1D MC simulations ($\blacktriangle$) with $N=50$ are shown
(in arbitrary units).
See text for details.
} 
\label{scaling_1D_fig}
\end{figure}

Also in 1D the translocation time obtained from LD simulations is
found to scale with the pore force as $\tau \sim f^{-1}$,
Fig.~\ref{scaling_1D_fig}~a). MC and KMC models also give this
scaling with pore force $f < 1$. However, for $f \ge 1 $ the
translocation time levels off to a constant value. Potentially, KMC
might not be as sensitive to this artifact, as will be discussed in
more detail later. However, KMC algorithm fails to run with so large a
force, due to the above-explained higher probability of moving a
particle regardless of whether a bond breaks or not. In 2D the
breaking of the force scaling has been reported in MC forced
translocation simulations in two~\cite{luo_mc} and three~\cite{loebl}
dimensions. Our simulations strongly imply that it is related to
the saturation of the transition probabilities in MC at large force values
inside the pore (see Appendix A, step 6).

The discrepancy between the different $\beta$ values obtained from MC
simulations can be explained by this artifact related to Metropolis
sampling. The infinite force for which $\beta = 1.53$ was
obtained~\cite{kantor} takes the MC model to the plateau region ($f >
1$), where $f=\infty$ is no different from $f=1$ in that it gives
identical velocity, $v$, to the translocating polymer, when the pore
length, $\Delta z$, is constant. However, by increasing $\Delta z$
while keeping the force acting per polymer bead, $f$, constant,
does increase pore potential and thus the polymer velocity, $v$, which
in turn, according to our simulations, increases $\beta$ (see
Fig.~\ref{scaling_moldy_fig} and Table~\ref{beta_values}). The
pore potential, $\Delta U$, over the pore is explicitly shown to
be the primary control parameter for forced translocation in
Table~\ref{porelength}, where translocation times are compared when
either the force acting on polymer beads inside the pore or the pore
length is changed in the LD model.  
\begin{table}[ht]
\caption{
Results from 3D Langevin dynamics for a polymer of length $N=100$.
Here $\Delta U \approx f \Delta z / k_B T$ is the dimensionless potential
over the pore.
}
\begin{tabular}{cccc}
$\Delta U$ & $f$ & $\Delta z$ & $\tau$ \\
\hline
30 & 10 & 3 & $326 \pm 10$ \\
15 & 10 & 1.5 & $642 \pm 20$ \\
30 & 20 & 1.5 & $354 \pm 10$ \\
\end{tabular}
\label{porelength}
\end{table}
In~\cite{kantor} $\Delta z = 1$, while in~\cite{luo_mc} $\Delta z
= 3$ was used. Of the two force values $f = 1, 5$ in~\cite{luo_mc} the
latter is within the plateau regime and the first at least very close
to it, which would explain the obtained almost identical $\beta$
values. Due to $f = \infty$ being in effect just the constant force
value in the plateau regime, $\Delta U$ in~\cite{kantor} is lower than
what was used in~\cite{luo_mc}, which explains why this $\beta$ value
for the infinite force is lower than what was obtained for force values $1$
and $5$. The saturation of the transition probabilities could also be the
reason why in 3D MC simulation results presented
in~\cite{dubbeldam_epl} to support FFE predictions the
value of $\beta$ did not change, when the pore force was varied. Hence,
it would seem that there is no reason to assume a universal $\beta$
for all $f$. 

The prevailing discrepancy between simulations can be at least partly
explained by the deficiency of the MC method at large force
regime, which is the biologically and experimentally relevant force
range. For example the Kuhn length of the  single-stranded DNA (ssDNA)
$\tilde{b} = 1.6$ nm, and a charge density $1.28\ e / \text{nm}$ reported
in~\cite{dessinges} for free ssDNA in an ionic solvent would result in the
effective charge of $2\ e/\tilde{b}$. Taking into account that the pore is
prone to screen the charges even more than the solvent~\cite{sauerbudge}
the effective charge on a nucleotide traversing the pore was estimated to be
$\approx 0.1\ e$.  In the $\alpha$-hemolysin pore of length $5.2$
nm there are approximately $13$ nucleotides of length
$0.4$ nm~\cite{meller_prl}, yielding total effective charge $q^* =
1.3\ e$ for the polymer segment inside the pore. Typical experimental
pore potential is $V \approx 120$
mV~\cite{kasianowicz,meller,storm,li}. At $T= 300 {\text K }$, the
dimensionless ratio $U = q^* V/k_B T \approx 6.03$ is obtained, 
which corresponds to the dimensionless value $U = \tilde{f} \Delta z / k_B
T$. Accordingly, the corresponding dimensionless force $f = \tilde{f}
b / k_B T$ has a value of $\simeq 2$ in our MC simulations, which is
already well in the plateau regime, see Fig.~\ref{scaling_1D_fig}~a).

We checked the effect of the pore friction in 1D, where dynamics
is no more constrained inside than outside the pore, making comparison
of the cases with frictional and frictionless pores
straightforward. The same characteristics as in higher dimensions were
seen. With a frictional pore $\beta=2$ is obtained only for $N \ge
30$. From 1D LD simulations with a frictional pore this
maximum length, $l_{max}$, was seen to increase with pore force,
$f$. Increasing $f$ increases the average velocity of the polymer,
$\langle v \rangle$, and so increases frictional contribution. Since
the force acts only on the part of the polymer inside the pore, at large
enough force frictional term damps movement only in the direction from
the {\it cis} to {\it trans} side thus increasing the translocation
time proportionately more for short than long polymers. At a very weak
force this effect is not perceived due to friction affecting movement
in both directions ({\it i.e.} also from {\it trans} to {\it cis}), as
seen also in dimensions 2 and 3 from the
topmost curves in Figs.~\ref{scaling_moldy_fig}~b) and c). This is by
definition a finite-size effect and bound to affect the obtained
$\beta$ more for shorter polymers, where the portion in the pore constitutes a
larger part. The decrease of $\beta$ due to pore friction is understandable,
since in the asymptotic limit of completely friction-dominated dynamics
linear scaling, $\beta = 1$, with polymer length is to be
expected.
\begin{figure}[htbp]
\centerline{
\includegraphics[angle=0,height=0.13\textheight]{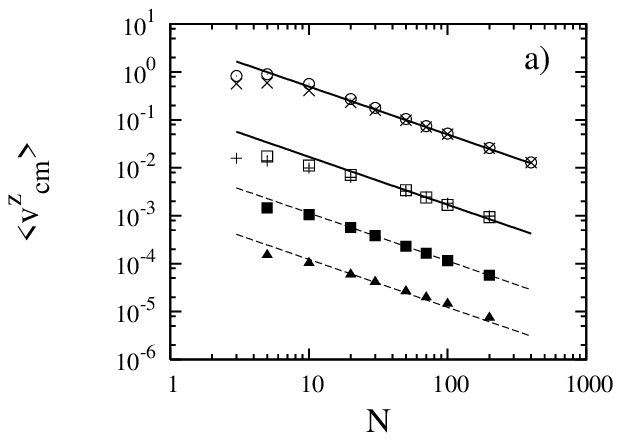}
\includegraphics[angle=0,height=0.13\textheight]{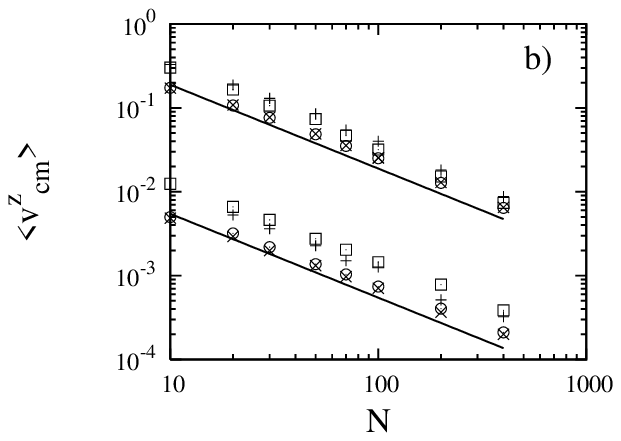}
}
\caption {
a) The average center-of-mass motion along the translocation
coordinate, $\langle v_{cm}^z \rangle$, as a function of the
chain length $N$ in 1D, with Langevin dynamics.
Two different pore forces $f=0.1$ ($+,\square$) and $20$ ($\times, \circ$)
have been used.
Both frictional ($+, \times$) and frictionless ($\square, \circ$) pores
were used.
MC results in 1D with $f=0.1$ ($\blacktriangle$) and $20$ ($\blacksquare$) are
also shown (in arbitrary units).
The solid and dashed lines have a slope of $-1$.
b) $\langle v_{cm}^z \rangle$ as a function of the
chain length $N$. Results are from Langevin dynamics simulations
in 2D ($\square, \circ$) and 3D ($+,\times$),
with two different pore forces $f=0.25,10$ for the lower and higher points,
respectively.
Both frictional ($+, \square$) and frictionless ($\times, \circ$)
pores were used. The solid lines $\sim 1/N$ are plotted to guide the eye.
}
\label{vel_Nscale}
\end{figure}

We measured the center-of-mass velocity, $\langle v_{cm}^z \rangle$, in the
$z$-direction, which is the direction perpendicular to the wall from {\it
cis} to {\it trans}, with a frictional and frictionless pore, see
Fig.~\ref{scaling_1D_fig}~b). Comparing the 1D curves obtained from LD
and MC in Figs.~\ref{scaling_1D_fig}~a) and b) it is clear that $\langle
v_{cm}^z \rangle$ is not simply reciprocal of $\tau$, which indicates that
one must be cautious in applying equilibrium concepts and observables in
characterizing translocation. Based on this finding, using $\langle z_0
\rangle / \tau$, where $z_0$ is the $z$ coordinate of the last polymer bead
in the initial equilibrated configuration, is not equivalent to using
$\langle v_{cm}^z \rangle$ for translocation velocity. In 1D it is
particularly clear, that the first definition simply sets the translocation
velocity $v \sim 1/\tau$, a valid definition for translocation
velocity, but one that also easily leads to drawing wrong
conclusions about translocation dynamics. Comparing
Fig.~\ref{scaling_moldy_fig}~a) and Fig.~\ref{scaling_1D_fig}~b), the same
characteristics is seen in dimensions 2 and 3. From
Fig.~\ref{scaling_1D_fig}~b) it is seen first of all that the artifact
in MC at large force magnitudes manifests itself in $\langle
v_{cm}^z \rangle$ measured in the 1D system. Outside of that, the $\langle
v_{cm}^z(f) \rangle$ curves for 1D MC and LD are qualitatively fairly
similar. The additional regime at very weak force magnitudes in LD was
checked to be related with the form of the FENE potential, showing
distinctively in 1D where it is the only potential between polymer beads.

To make sure that we are seeing finite-size effects for short polymers
with a frictional pore instead of two scaling regimes for short and
long polymers, we also check the dependence of $\langle
v_{cm}^z(f) \rangle$ on polymer length, $N$. In
accord with the findings in~\cite{kantor} the center-of-mass motion of
a polymer translocating through a pore closely conforms to {\it unimpeded
motion}, where $\langle v_{cm}^z \rangle \sim 1/N$ with both the
frictional and frictionless pore, see Fig.~\ref{vel_Nscale}. It is
evident that pore friction does not affect the behavior of $\langle
v_{cm}^z \rangle$ that is a measure of the transfer of the whole
polymer. Hence, the differences in Figs.~\ref{scaling_moldy_fig}~b)
and c) have to come solely from the
frictional contribution to the finite portion of the chain in the
pore. The effect is seen clearly in $\tau$ that is a local observable
quantifying the transfer of the segments over the pore.

\begin{figure}
\centerline{
\includegraphics[angle=0, width=0.3\textwidth]{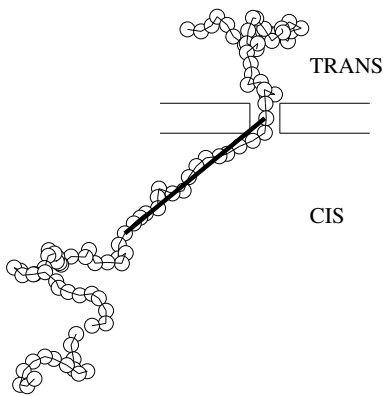}
}
\caption{
A snapshot at $s=35$ of a 3D LD simulation with a polymer of length $N=100$.
The dimensionless pore force used here is $f=0.01$. The wide solid line
represents the pore-to-end distance $R_{pe}$ as the number of mobile beads,
$s_m$, reaches the index of the observed bead, $n=52$.
}
\label{snapshot_conf}
\end{figure}
\begin{figure}[!ht]
\centerline{
\includegraphics[angle=0, height=0.13\textheight]{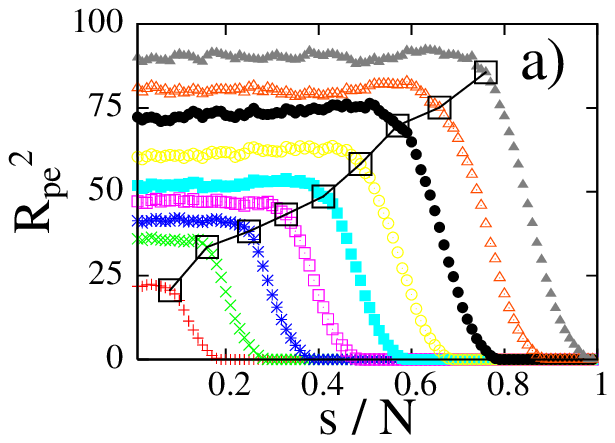}
\includegraphics[angle=0, height=0.13\textheight]{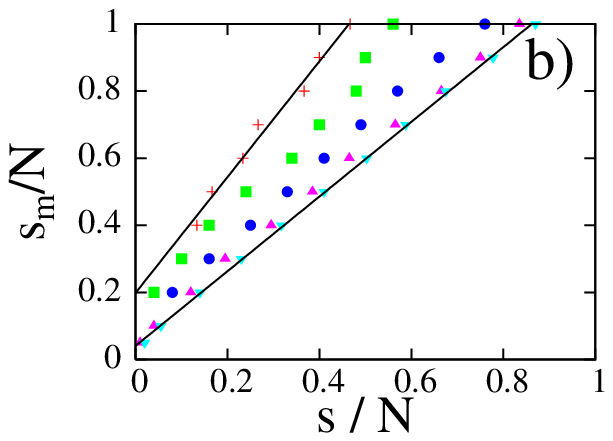}
}
\caption[Ree] {
  (Color online)
  a) 3D Langevin dynamics. Averaged squared distances of beads
  numbered $30,50,70,$, $100$, $200$, and $400$ from the pore as a
  function of the number of translocated beads $s$ for polymers of
  length $N = 100$. The force $f = 0.25$. 
  b) The number of mobile beads, $s_m$ vs the number of translocated beads,
  $s$, both normalized to the polymer length, $N$.
}
\label{Ree}
\end{figure}
\begin{figure}
\centerline{
\includegraphics[angle=0, height=0.25\textheight]{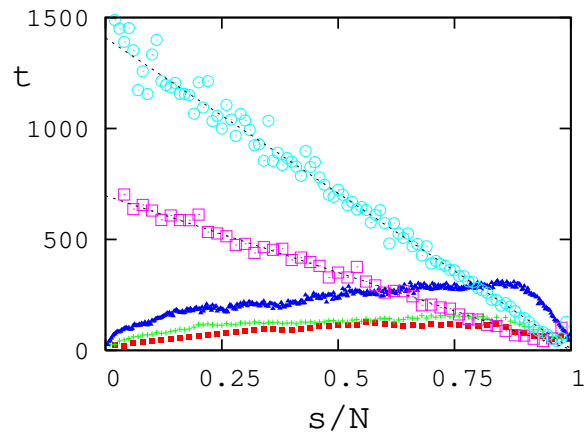}
}
\caption{
(Color online) 3D LD. Waiting time averages $t(s)$ as a function of the
translocated segment $s$ with $f/D_0 = 2.91$.
Profiles with filled ($N=50$: $\blacksquare$, $N=100$: $\bullet$,
$N=400$: $\blacktriangle$) and empty symbols ($N=50$: $\square$,
$N=100$: $\circ$) represent runs with an equilibrium initial
configuration and with a straight initial configuration,
respectively. Note that the latter profiles yield a clear linear behavior.
}
\label{time_averages}
\end{figure}

The above observation that $\langle v_{cm}^z(f) \rangle \nsim
1/\tau(f)$ indicated that the transfer of the part of the polymer
crucial for characterization of forced translocation, {\it i.e.} the
part in the vicinity of the pore does not remain even close to thermal
equilibrium. Any meaningful description of the translocation process must
then succeed to capture the mechanics of this part. Given that any
diffusive, or close-to-equilibrium description is invalid, the
remaining minimum condition is force balance. Indeed, it turns out
that the observed scaling behavior $\beta \leq 1 + \nu$ in 2D and 3D
can be explained by the force balance of the drag force exerted on the
mobile beads on the {\it cis} side and the constant pore force, $f_d =
f$. Just as in~\cite{lehtola} we measured the squared distance, $R_{pe}^2(n)$
(see Figs.~\ref{straight_conf}~b) and \ref{snapshot_conf}), of
the polymer bead $n$ from the pore on the {\it cis} side during
translocation, see Fig.~\ref{Ree}~a), from which the segments toward
the free end of the polymer are seen to remain immobile until they are
pulled toward the pore. Fig.~\ref{Ree}~b) shows that the number of mobile beads
increases linearly as a function of translocated beads, $s_m = k
s$. Up to lengths of $N \approx 100$, $k \sim N^{-\chi}$, beyond which
it gradually levels off to a constant value greater than unity. 
The mobile beads comprise a moving segment, which already for moderate
pore force exhibit no folding indicating that the relaxation of
the beads in the moving segment is far slower than the rate at which
they are translocated, see Figs.~\ref{straight_conf}~c) and
\ref{snapshot_conf}. $f_d$ is exerted
on the mobile beads, so $f_d \sim s_m \langle v \rangle$, where $\langle v
\rangle$ is the average velocity of the mobile beads. When the whole
chain has translocated, $f_d \sim N_m \langle v \rangle$, where $N_m =
k N$. The beads are set in motion from their equilibrium positions, so
the distance $d$ of the last bead to be translocated scales as
$d \sim N^\nu$. The average translocation time then scales as $\tau
\sim \langle d \rangle / \langle v \rangle \sim k N^{1+\nu} \sim
N^{1+\nu-\chi}$. For example for $f/D_0 = 2.9$, we obtain $\chi
\approx 0.3$ from the data displayed in Fig. \ref{Ree}~b),
which would give $\beta \approx 1.3$. The crowding of polymer beads close to
the pore on the {\it trans} side (see Fig.~\ref{straight_conf}~c)
increases $\beta$ from this value. The crowding increases with force,
which has been shown in~\cite{lehtola}. From Fig.~\ref{Ree}~b)
it can be seen that $k$ saturates when $f$ is increased, which also
increases $\beta$ at large $f$.

In order to confirm that the presence of $\nu$ in the scaling exponent
$\beta = 1 + \nu$ comes solely from the initial configuration and not
from any factor due to close-to-equilibrium motion and Rouse
relaxation time, we check the scaling and the average waiting time, $t(s)$, for
the segment $s$ to translocate for an initially extended
configuration, Fig.~\ref{straight_conf}~a). We obtain $\beta = 2$ for
this initial configuration. The average waiting time profiles of
the initially extended and equilibrated configurations are shown in
Fig.~\ref{time_averages}. For the initially extended configuration we
obtain linear dependence $t(s) \sim s$ giving exactly the obtained scaling with
$\beta=2$.  Hence, the upper limit for the scaling exponent $\beta$
results from the waiting times of the translocating beads, determined 
solely on their initial positions. In other words, already with a
moderate pore force translocation velocity completely dominates the
process so that diffusive motion has no effect and the the only
prevailing condition governing translocation dynamics is the force balance
described above.

\section{Summary}

We have studied forced translocation without hydrodynamics in
different dimensions using Monte Carlo (MC), kinetic Monte Carlo
(KMC), and Langevin dynamics (LD) methods. We have shown that forced
translocation model using MC with basic Metropolis sampling gives the
correct time dependence at moderate pore potentials but presents an
artifact at large pore potentials. This artifact seems to explain the
prevailing controversy between different MC results. It also seems to
account for results claiming universal scaling of the translocation
time with the polymer length, $\tau \sim N^\beta$, independent of the
pore force magnitude and hence the prevailing discrepancy between 
computational and experimental findings on forced translocation.

Using LD in 2D and 3D we have shown that the scaling exponent $\beta$
increases with the pore force, $f$. We obtain $\beta = 1+ \nu$ as the
high force limit. By measurements of polymers'
center-of-mass velocities we have shown that description of forced
translocation with concepts related to close-to-equilibrium
concepts is not well founded. We have given a description of the
forced translocation based on simple force-balance condition, where
the drag force exerted on the part of the polymer on the {\it cis}
side changes with the number of polymer beads in motion. This
description gives the above upper limit for $\beta$ and also explains
the increase of $\beta$ with $f$. We have also shown that crowding of
polymer beads is strong in the vicinity of the pore on the {\it trans} side.

This work has been supported by the Academy of Finland (Project No.~127766).


\bibliography{times}

\begin{thebibliography}{30}
\expandafter\ifx\csname natexlab\endcsname\relax\def\natexlab#1{#1}\fi
\expandafter\ifx\csname bibnamefont\endcsname\relax
  \def\bibnamefont#1{#1}\fi
\expandafter\ifx\csname bibfnamefont\endcsname\relax
  \def\bibfnamefont#1{#1}\fi
\expandafter\ifx\csname citenamefont\endcsname\relax
  \def\citenamefont#1{#1}\fi
\expandafter\ifx\csname url\endcsname\relax
  \def\url#1{\texttt{#1}}\fi
\expandafter\ifx\csname urlprefix\endcsname\relax\def\urlprefix{URL }\fi
\providecommand{\bibinfo}[2]{#2}
\providecommand{\eprint}[2][]{\url{#2}}

\bibitem[{\citenamefont{Ali and Yeomans}(2005)}]{ali}
\bibinfo{author}{\bibfnamefont{I.}~\bibnamefont{Ali}} \bibnamefont{and}
  \bibinfo{author}{\bibfnamefont{J.~M.} \bibnamefont{Yeomans}},
  \bibinfo{journal}{J. Chem. Phys.} \textbf{\bibinfo{volume}{123}},
  \bibinfo{pages}{234903} (\bibinfo{year}{2005}).

\bibitem[{\citenamefont{Fyta et~al.}(2008)\citenamefont{Fyta, Kaxiras,
  Melchionna, and Succi}}]{fyta}
\bibinfo{author}{\bibfnamefont{M.}~\bibnamefont{Fyta}},
  \bibinfo{author}{\bibfnamefont{E.}~\bibnamefont{Kaxiras}},
  \bibinfo{author}{\bibfnamefont{S.}~\bibnamefont{Melchionna}},
  \bibnamefont{and} \bibinfo{author}{\bibfnamefont{S.}~\bibnamefont{Succi}},
  \bibinfo{journal}{Comp. Sci. \& Eng.} \textbf{\bibinfo{volume}{2}},
  \bibinfo{pages}{20} (\bibinfo{year}{2008}).

\bibitem[{\citenamefont{Bernaschi et~al.}(2008)\citenamefont{Bernaschi,
  Melchionna, Succi, Fyta, and Kaxiras}}]{bernaschi}
\bibinfo{author}{\bibfnamefont{M.}~\bibnamefont{Bernaschi}},
  \bibinfo{author}{\bibfnamefont{S.}~\bibnamefont{Melchionna}},
  \bibinfo{author}{\bibfnamefont{S.}~\bibnamefont{Succi}},
  \bibinfo{author}{\bibfnamefont{M.}~\bibnamefont{Fyta}}, \bibnamefont{and}
  \bibinfo{author}{\bibfnamefont{E.}~\bibnamefont{Kaxiras}},
  \bibinfo{journal}{Nanoletters} \textbf{\bibinfo{volume}{8}},
  \bibinfo{pages}{1115} (\bibinfo{year}{2008}).

\bibitem[{\citenamefont{Gauthier and Slater}(2008{\natexlab{a}})}]{gauthier}
\bibinfo{author}{\bibfnamefont{M.~G.} \bibnamefont{Gauthier}} \bibnamefont{and}
  \bibinfo{author}{\bibfnamefont{G.~W.} \bibnamefont{Slater}},
  \bibinfo{journal}{Eur. Phys. J. E} \textbf{\bibinfo{volume}{25}},
  \bibinfo{pages}{17} (\bibinfo{year}{2008}{\natexlab{a}}).

\bibitem[{\citenamefont{Lehtola et~al.}(2008)\citenamefont{Lehtola, Linna, and
  Kaski}}]{lehtola}
\bibinfo{author}{\bibfnamefont{V.~V.}~\bibnamefont{Lehtola}},
  \bibinfo{author}{\bibfnamefont{R.~P.} \bibnamefont{Linna}}, \bibnamefont{and}
  \bibinfo{author}{\bibfnamefont{K.}~\bibnamefont{Kaski}},
  \bibinfo{journal}{(submitted to Phys. Rev. Lett.)}  (\bibinfo{year}{2008}).

\bibitem[{\citenamefont{Sung and Park}(1996)}]{sung}
\bibinfo{author}{\bibfnamefont{W.}~\bibnamefont{Sung}} \bibnamefont{and}
  \bibinfo{author}{\bibfnamefont{P.~J.} \bibnamefont{Park}},
  \bibinfo{journal}{Phys. Rev. Lett.} \textbf{\bibinfo{volume}{77}},
  \bibinfo{pages}{783} (\bibinfo{year}{1996}).

\bibitem[{\citenamefont{Muthukumar}(1999)}]{muthukumar}
\bibinfo{author}{\bibfnamefont{M.}~\bibnamefont{Muthukumar}},
  \bibinfo{journal}{J. Chem. Phys.} \textbf{\bibinfo{volume}{111}},
  \bibinfo{pages}{10371} (\bibinfo{year}{1999}).

\bibitem[{\citenamefont{Chuang et~al.}(2001)\citenamefont{Chuang, Kantor, and
  Kardar}}]{chuang}
\bibinfo{author}{\bibfnamefont{J.}~\bibnamefont{Chuang}},
  \bibinfo{author}{\bibfnamefont{Y.}~\bibnamefont{Kantor}}, \bibnamefont{and}
  \bibinfo{author}{\bibfnamefont{M.}~\bibnamefont{Kardar}},
  \bibinfo{journal}{Phys. Rev. E} \textbf{\bibinfo{volume}{65}},
  \bibinfo{pages}{011802} (\bibinfo{year}{2001}).

\bibitem[{\citenamefont{Kantor and Kardar}(2004)}]{kantor}
\bibinfo{author}{\bibfnamefont{Y.}~\bibnamefont{Kantor}} \bibnamefont{and}
  \bibinfo{author}{\bibfnamefont{M.}~\bibnamefont{Kardar}},
  \bibinfo{journal}{Phys. Rev. E} \textbf{\bibinfo{volume}{69}},
  \bibinfo{pages}{021806} (\bibinfo{year}{2004}).

\bibitem[{\citenamefont{Dubbeldam
  et~al.}(2007{\natexlab{a}})\citenamefont{Dubbeldam, Milchev, Rostiashvili,
  and Vilgis}}]{dubbeldam_pre}
\bibinfo{author}{\bibfnamefont{J.~L.~A.} \bibnamefont{Dubbeldam}},
  \bibinfo{author}{\bibfnamefont{A.}~\bibnamefont{Milchev}},
  \bibinfo{author}{\bibfnamefont{V.~G.} \bibnamefont{Rostiashvili}},
  \bibnamefont{and} \bibinfo{author}{\bibfnamefont{T.~A.}
  \bibnamefont{Vilgis}}, \bibinfo{journal}{Phys. Rev. E}
  \textbf{\bibinfo{volume}{76}}, \bibinfo{pages}{010801(R)}
  (\bibinfo{year}{2007}{\natexlab{a}}).

\bibitem[{\citenamefont{Dubbeldam
  et~al.}(2007{\natexlab{b}})\citenamefont{Dubbeldam, Milchev, Rostiashvili,
  and Vilgis}}]{dubbeldam_epl}
\bibinfo{author}{\bibfnamefont{J.~L.~A.} \bibnamefont{Dubbeldam}},
  \bibinfo{author}{\bibfnamefont{A.}~\bibnamefont{Milchev}},
  \bibinfo{author}{\bibfnamefont{V.~G.} \bibnamefont{Rostiashvili}},
  \bibnamefont{and} \bibinfo{author}{\bibfnamefont{T.~A.}
  \bibnamefont{Vilgis}}, \bibinfo{journal}{Euro. Phys. Lett.}
  \textbf{\bibinfo{volume}{79}}, \bibinfo{pages}{18002}
  (\bibinfo{year}{2007}{\natexlab{b}}).

\bibitem[{\citenamefont{Luo et~al.}(2006)\citenamefont{Luo, Huopaniemi,
  Ala-Nissila, and Ying}}]{luo_mc}
\bibinfo{author}{\bibfnamefont{K.}~\bibnamefont{Luo}},
  \bibinfo{author}{\bibfnamefont{I.}~\bibnamefont{Huopaniemi}},
  \bibinfo{author}{\bibfnamefont{T.}~\bibnamefont{Ala-Nissila}},
  \bibnamefont{and} \bibinfo{author}{\bibfnamefont{S.-C.} \bibnamefont{Ying}},
  \bibinfo{journal}{J. Chem. Phys.} \textbf{\bibinfo{volume}{124}},
  \bibinfo{pages}{114704} (\bibinfo{year}{2006}).

\bibitem[{\citenamefont{Huopaniemi et~al.}(2006)\citenamefont{Huopaniemi, Luo,
  Ala-Nissila, and Ying}}]{luo_langevin}
\bibinfo{author}{\bibfnamefont{I.}~\bibnamefont{Huopaniemi}},
  \bibinfo{author}{\bibfnamefont{K.}~\bibnamefont{Luo}},
  \bibinfo{author}{\bibfnamefont{T.}~\bibnamefont{Ala-Nissila}},
  \bibnamefont{and} \bibinfo{author}{\bibfnamefont{S.-C.} \bibnamefont{Ying}},
  \bibinfo{journal}{J. Chem. Phys.} \textbf{\bibinfo{volume}{125}},
  \bibinfo{pages}{124901} (\bibinfo{year}{2006}).

\bibitem[{\citenamefont{Gauthier and
  Slater}(2008{\natexlab{b}})}]{gauthier_jcp}
\bibinfo{author}{\bibfnamefont{M.~G.} \bibnamefont{Gauthier}} \bibnamefont{and}
  \bibinfo{author}{\bibfnamefont{G.~W.} \bibnamefont{Slater}},
  \bibinfo{journal}{J. Chem. Phys.} \textbf{\bibinfo{volume}{128}},
  \bibinfo{pages}{205103} (\bibinfo{year}{2008}{\natexlab{b}}).

\bibitem[{\citenamefont{Tian and Smith}(2003)}]{tian}
\bibinfo{author}{\bibfnamefont{P.}~\bibnamefont{Tian}} \bibnamefont{and}
  \bibinfo{author}{\bibfnamefont{G.~D.} \bibnamefont{Smith}},
  \bibinfo{journal}{J. Chem. Phys.} \textbf{\bibinfo{volume}{119}},
  \bibinfo{pages}{11475} (\bibinfo{year}{2003}).

\bibitem[{\citenamefont{Loebl et~al.}(2003)\citenamefont{Loebl, Randel,
  Goodwin, and Matthai}}]{loebl}
\bibinfo{author}{\bibfnamefont{H.~C.} \bibnamefont{Loebl}},
  \bibinfo{author}{\bibfnamefont{R.}~\bibnamefont{Randel}},
  \bibinfo{author}{\bibfnamefont{S.~P.} \bibnamefont{Goodwin}},
  \bibnamefont{and} \bibinfo{author}{\bibfnamefont{C.~C.}
  \bibnamefont{Matthai}}, \bibinfo{journal}{Phys. Rev. E}
  \textbf{\bibinfo{volume}{67}}, \bibinfo{pages}{041913}
  (\bibinfo{year}{2003}).

\bibitem[{\citenamefont{Forrey and Muthukumar}(2007)}]{forrey}
\bibinfo{author}{\bibfnamefont{C.}~\bibnamefont{Forrey}} \bibnamefont{and}
  \bibinfo{author}{\bibfnamefont{M.}~\bibnamefont{Muthukumar}},
  \bibinfo{journal}{J. Chem. Phys.} \textbf{\bibinfo{volume}{127}},
  \bibinfo{pages}{015102} (\bibinfo{year}{2007}).

\bibitem[{\citenamefont{Storm~{\it et al}}(2005)}]{storm}
\bibinfo{author}{\bibfnamefont{W.}~\bibnamefont{Storm~{\it et al}}},
  \bibinfo{journal}{Nano Lett.} \textbf{\bibinfo{volume}{5}},
  \bibinfo{pages}{1193} (\bibinfo{year}{2005}).

\bibitem[{\citenamefont{Kasianowicz et~al.}(1996)\citenamefont{Kasianowicz,
  Brandin, Branton, and Deamer}}]{kasianowicz}
\bibinfo{author}{\bibfnamefont{J.~J.} \bibnamefont{Kasianowicz}},
  \bibinfo{author}{\bibfnamefont{E.}~\bibnamefont{Brandin}},
  \bibinfo{author}{\bibfnamefont{D.}~\bibnamefont{Branton}}, \bibnamefont{and}
  \bibinfo{author}{\bibfnamefont{D.~W.} \bibnamefont{Deamer}},
  \bibinfo{journal}{Proc. Natl. Acad. Sci. U.S.A.}
  \textbf{\bibinfo{volume}{93}}, \bibinfo{pages}{13770} (\bibinfo{year}{1996}).

\bibitem[{\citenamefont{Meller}(2003)}]{meller}
\bibinfo{author}{\bibfnamefont{A.}~\bibnamefont{Meller}}, \bibinfo{journal}{J.
  Phys. Condens. Matter} \textbf{\bibinfo{volume}{15}}, \bibinfo{pages}{R581}
  (\bibinfo{year}{2003}).

\bibitem[{\citenamefont{Li~{\it et al.}}(2003)}]{li}
\bibinfo{author}{\bibfnamefont{J.}~\bibnamefont{Li~{\it et al.}}},
  \bibinfo{journal}{Nature Materials} \textbf{\bibinfo{volume}{2}},
  \bibinfo{pages}{611} (\bibinfo{year}{2003}).

\bibitem[{\citenamefont{Chen}(1988)}]{chen}
\bibinfo{author}{\bibfnamefont{L.~B.} \bibnamefont{Chen}},
  \bibinfo{journal}{Ann. Rev. Cell Biol.} \textbf{\bibinfo{volume}{4}},
  \bibinfo{pages}{155} (\bibinfo{year}{1988}).

\bibitem[{\citenamefont{Huang et~al.}(2002)\citenamefont{Huang, Ratliff, and
  Matouschek}}]{huang}
\bibinfo{author}{\bibfnamefont{S.}~\bibnamefont{Huang}},
  \bibinfo{author}{\bibfnamefont{K.~S.} \bibnamefont{Ratliff}},
  \bibnamefont{and}
  \bibinfo{author}{\bibfnamefont{A.}~\bibnamefont{Matouschek}},
  \bibinfo{journal}{Nature Struct. Biol.} \textbf{\bibinfo{volume}{9}},
  \bibinfo{pages}{301} (\bibinfo{year}{2002}).

\bibitem[{\citenamefont{Young and Elcock}(1966)}]{young}
\bibinfo{author}{\bibfnamefont{W.~M.} \bibnamefont{Young}} \bibnamefont{and}
  \bibinfo{author}{\bibfnamefont{E.~W.} \bibnamefont{Elcock}},
  \bibinfo{journal}{Proc. Phys. Soc.} \textbf{\bibinfo{volume}{89}},
  \bibinfo{pages}{735} (\bibinfo{year}{1966}).

\bibitem[{\citenamefont{Bortz et~al.}(1975)\citenamefont{Bortz, Kalos, and
  Lebowitz}}]{bortz}
\bibinfo{author}{\bibfnamefont{A.~B.} \bibnamefont{Bortz}},
  \bibinfo{author}{\bibfnamefont{M.~H.} \bibnamefont{Kalos}}, \bibnamefont{and}
  \bibinfo{author}{\bibfnamefont{J.~L.} \bibnamefont{Lebowitz}},
  \bibinfo{journal}{J. Comp. Phys.} \textbf{\bibinfo{volume}{17}},
  \bibinfo{pages}{10} (\bibinfo{year}{1975}).

\bibitem[{\citenamefont{Fichthorn and Weinberg}(1991)}]{fichthorn}
\bibinfo{author}{\bibfnamefont{K.~A.} \bibnamefont{Fichthorn}}
  \bibnamefont{and} \bibinfo{author}{\bibfnamefont{W.~H.}
  \bibnamefont{Weinberg}}, \bibinfo{journal}{J. Chem. Phys.}
  \textbf{\bibinfo{volume}{95}}, \bibinfo{pages}{1090} (\bibinfo{year}{1991}).

\bibitem[{\citenamefont{Allen and Tildesley}(2006)}]{allen}
\bibinfo{author}{\bibfnamefont{M.}~\bibnamefont{Allen}} \bibnamefont{and}
  \bibinfo{author}{\bibfnamefont{D.~J.} \bibnamefont{Tildesley}},
  \emph{\bibinfo{title}{Computer Simulation of Liquids}}
  (\bibinfo{publisher}{Oxford science publications}, \bibinfo{address}{Oxford},
  \bibinfo{year}{2006}).

\bibitem[{\citenamefont{Dessinges et~al.}(2002)\citenamefont{Dessinges, Maier,
  Zhang, Peliti, Bensimon, and Croquette}}]{dessinges}
\bibinfo{author}{\bibfnamefont{M.-N.} \bibnamefont{Dessinges}},
  \bibinfo{author}{\bibfnamefont{B.}~\bibnamefont{Maier}},
  \bibinfo{author}{\bibfnamefont{Y.}~\bibnamefont{Zhang}},
  \bibinfo{author}{\bibfnamefont{M.}~\bibnamefont{Peliti}},
  \bibinfo{author}{\bibfnamefont{D.}~\bibnamefont{Bensimon}}, \bibnamefont{and}
  \bibinfo{author}{\bibfnamefont{V.}~\bibnamefont{Croquette}},
  \bibinfo{journal}{Phys. Rev. Lett.} \textbf{\bibinfo{volume}{89}},
  \bibinfo{pages}{248102} (\bibinfo{year}{2002}).

\bibitem[{\citenamefont{Sauer-Budge et~al.}(2003)\citenamefont{Sauer-Budge,
  Nyamwanda, Lubensky, and Branton}}]{sauerbudge}
\bibinfo{author}{\bibfnamefont{A.~F.} \bibnamefont{Sauer-Budge}},
  \bibinfo{author}{\bibfnamefont{J.~A.} \bibnamefont{Nyamwanda}},
  \bibinfo{author}{\bibfnamefont{D.~K.} \bibnamefont{Lubensky}},
  \bibnamefont{and} \bibinfo{author}{\bibfnamefont{D.}~\bibnamefont{Branton}},
  \bibinfo{journal}{Phys. Rev. Lett.} \textbf{\bibinfo{volume}{90}},
  \bibinfo{pages}{238101} (\bibinfo{year}{2003}).

\bibitem[{\citenamefont{Meller et~al.}(2001)\citenamefont{Meller, Nivon, and
  Branton}}]{meller_prl}
\bibinfo{author}{\bibfnamefont{A.}~\bibnamefont{Meller}},
  \bibinfo{author}{\bibfnamefont{L.}~\bibnamefont{Nivon}}, \bibnamefont{and}
  \bibinfo{author}{\bibfnamefont{D.}~\bibnamefont{Branton}},
  \bibinfo{journal}{Phys. Rev. Lett.} \textbf{\bibinfo{volume}{86}},
  \bibinfo{pages}{3435} (\bibinfo{year}{2001}).

\end{thebibliography}

\section{Appendix A: Metropolis Monte Carlo algorithm}

\begin{enumerate}
\item Choose an initial state, and set the time $t = 0$.

\item Randomly choose a particle with label $i$, and calculate a trial position
$\vec{r'_i} = \vec{r_i} + \vec{\delta r_i}$.

\item Calculate the energy change, $\Delta U = -\vec{f_i} \cdot
\frac{\vec{\delta r_i}}{\lvert \vec{\delta r_i} \rvert} +
U_{FENE}(\vec{r_i}+\vec{\delta r_i}) - U_{FENE}(\vec{r_i})$, resulting
from this displacement.

\item If $\Delta U < 0$, the move is accepted; go to step 2.

\item Get a uniform random number $u \in (0, 1]$.
\item If $u < \exp(-\Delta U / k_B T)$, accept the move.
\item Update the time $t = t + 1/N$ and go to step 2.

\end{enumerate}

\section{Appendix B: Kinetic Monte Carlo Algorithm}

The KMC algorithm for simulating the time evolution of a system where
some processes occur with known average rates, or probabilities, $p_i
= \exp(-\Delta U)$, where $\Delta U$ is as defined in Appendix A,
can be written as follows:
\begin{enumerate}
\item Choose an initial state, set the time $t = 0$, and
form a list of all possible rates in the system $p_i$.
\item Calculate the cumulative function $R_i=\sum_{j=1}^i p_j$ for
 $i=1,\ldots,l$, where $l$ is the total number of transitions. Denote $R = R_l$.
\item Get a uniform random number $u \in (0, 1]$.
\item Find the event to carry out by finding the $i$ for which $R_{i-1} < u R < R_i$.
\item Carry out event $i$.
\item Recalculate all rates $p_i$ which may have changed due to the
  transition. If appropriate, remove or add new transitions
  $i$. Update $N$ and the list of event rates accordingly.
\item Get a new uniform random number $u' \in (0, 1]$.
\item Update the time with $t = t + \Delta t$, where 
$\Delta t = - { \frac{\log u}{R}}$.
\item Return to step 2.
\end{enumerate}
(note that the same average time scale can be obtained also using 
$\Delta t = \frac{1}{R}$ in step 8. However, including the random
number describes better the stochastic nature of the process).

This algorithm is known in different sources variously as the
residence-time algorithm or 
the n-fold way or the Bortz-Kalos-Lebowitz (BKL) algorithm or just the kinetic 
Monte Carlo (KMC) algorithm.

\end{document}